\documentclass[
 reprint,
 amsmath,amssymb,
 aps,
]{revtex4-2}

\usepackage{graphicx}
\usepackage{dcolumn}
\usepackage{bm}
\usepackage{physics,multirow,booktabs}

\begin{document}

\title{Modular Superconducting Qubit Architecture\\with a Multi-chip Tunable Coupler}

\author{Mark Field}\thanks{These authors contributed equally}
\author{Angela Q. Chen}\thanks{These authors contributed equally}
\author{Ben Scharmann}
\author{Eyob A. Sete}
\author{Feyza Oruc}
\author{\\Kim Vu}
\author{Valentin Kosenko}
\author{Joshua Y. Mutus}
\author{Stefano Poletto}
\author{Andrew Bestwick}\thanks{email andrew@rigetti.com}

\affiliation{
Rigetti Computing, 775 Heinz Avenue, Berkeley, CA 94710
}

\begin{abstract}
We use a floating tunable coupler to mediate interactions between qubits on separate chips to build a modular architecture. We demonstrate three different designs of multi-chip tunable couplers using vacuum gap capacitors or superconducting indium bump bonds to connect the coupler to a microwave line on a common substrate and then connect to the qubit on the next chip. We show that the zero-coupling condition between qubits on separate chips can be achieved in each design and that the relaxation rates for the coupler and qubits are not noticeably affected by the extra circuit elements. Finally, we demonstrate two-qubit gate operations with fidelity at the same level as qubits with a tunable coupler on a single chip. Using one or more indium bonds does not degrade qubit coherence or impact the performance of two-qubit gates.
\end{abstract}

\maketitle

\section{Introduction}
To realize the promise of quantum computing, we require systems with sufficiently low error rates to execute algorithms without being overwhelmed by noise and errors. For superconducting quantum processors, one of the challenges is being able to scale up the number of qubits in an architecture without introducing additional channels for correlated gate errors. One possible solution is to use a modular approach where small high-yielding quantum processors are assembled into larger systems using quantum coherent interconnects \cite{awschalom2021, jiang2007, kimble2008, chou2018,wan2019,hensen2015, monroe2014}. Ultimately, scaling quantum processors while maintaining gate errors below a threshold value should allow the error rate to be reduced by the use of error correction of logical qubits made up of many physical qubits~\cite{KITAEV20032,Fowler2012,Raussendorf2007,Krinner2022,Acharya2023}. 

On superconducting quantum processors, modular architectures allow larger devices to be built from smaller units, which increases yield and allows for pre-screening before integration. Such modular architectures enable better-performing devices for the assembled processor when compared with larger monolithic devices at the expense of more complex multi-chip device assembly \cite{brink2018,dickel2018}. Fabricating Josephson junctions to target qubit frequencies within system tolerances is a hard engineering task, with variation of resistances of order 2$\%$ within a chip \cite{kreikebaum2020}. Much better frequency targeting is possible on smaller die compared with larger die with more qubits. Modular devices also have the advantage of increased isolation between units reducing cross-talk and correlated errors, such as those produced by high-energy background radiation \cite{wilen2021,vepsalainen2020,cardani2021}. High-fidelity entanglement of qubits on separate chips was previously demonstrated using static capacitive coupling of frequency tunable qubits to a co-planar transmission line that spanned adjacent chips~\cite{gold2021}.

Suppressing error rates has been an active area of research to improve the performance of superconducting quantum processors. In particular, tunable couplers have been introduced as useful components to both minimize always-on ZZ interactions and implement fast entangling gates with two-qubit gate fidelities approaching 99.9\% \cite{yan2018,sung2021,foxen2020,collodo2020a,xu2020,arute2019a,mundada2019}. While tunable couplers have been investigated only on single-chip configurations, a direct qubit-qubit capacitance is not required for the floating tunable coupler, making it a potentially compatible component for modular architectures. 

In this work, we implement a floating tunable coupler \cite{sete2021,stehlik2021,marxer2023,liang2023} to couple qubits on separate chips to enable high performance in a scalable modular architecture. We first engineer a multi-chip tunable coupler design that is compatible with a modular architecture and robust against modular packaging complexities. Once we have this basic design we then consider three specific variations with increasing packaging complexity and experimentally validate that the multi-chip coupler designs can mediate both high- and low-coupling interactions between qubits on separate chips without degrading qubit coherence. We also demonstrate a 56 ns parametric-resonance CZ gate with fidelity as high as $99.13\%$ between qubits on separate chips, reaching performance that is comparable to the two-qubit gate realized on qubits fabricated on the same chip.

\section{Device Design and Fabrication}
\subsection{Design of a Floating Tunable Coupler} \label{design}

\begin{figure}[!htb]
    \centering
    \includegraphics[width=\columnwidth]{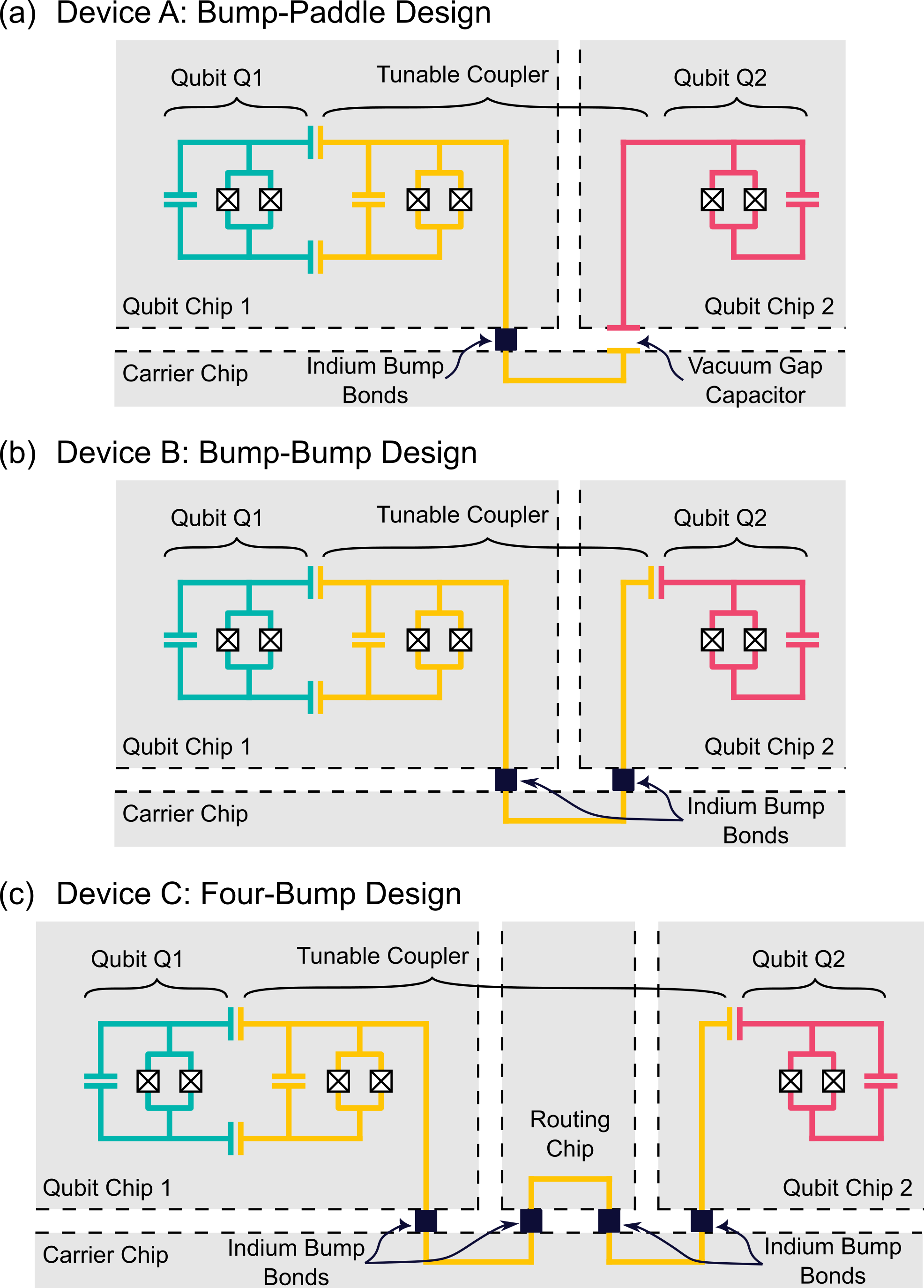}
    \caption{Circuit diagrams of the measured multi-chip coupler devices. The architecture consists of two separate qubit chips that are bonded to a common carrier chip. The floating multi-chip coupler (yellow) is defined on the same chip as the qubit labeled Q1 (teal), and one coupler arm extends across the carrier chip to the qubit labeled Q2 (magenta) on an adjacent chip. (a) On the Bump-Paddle design, the tunable coupler arm bridges the two chips via an indium bump connector on one side and a vacuum gap capacitor on the other side. The vacuum gap capacitor separation is set by the bump connector height. (b) An additional indium bump connector replaces the vacuum gap capacitor on the Bump-Bump design, so the tunable coupler arm spans two indium bumps. (c) An additional routing chip can be added between Qubit Chips 1 and 2 depending on assembly needs. In this Four-Bump design, the tunable coupler spans four surfaces (the two distinct qubit chips, the carrier chip which carries two separate signal lines, and the routing chip) using four indium bumps.}
    \label{fig:device diagram}
\end{figure}

Following Ref.~\cite{sete2021}, for two qubits coupled by a floating tunable coupler, the qubit-coupler and qubit-qubit couplings are given by:

\begin{align}
    g_{kc} &= \frac{E_{kc}}{\sqrt{2}}\left(\frac{E_{Jk}}{ E_{Ck}} \frac{E_{Jc}}{E_{Cc}}\right)^{\frac{1}{4}},~~k\in\{1,2\},\label{virtual}\\
    g_{12} &= \frac{E_{12}}{\sqrt{2}}\left(\frac{E_{J1}}{ E_{C1}} \frac{E_{J2}}{E_{C2}}\right)^{\frac{1}{4}},\label{g_direct}
\end{align}

where $E_{Jk}$ and $E_{Ck}$ are the Josephson and charging energies for the qubits $(k=1,2)$ and coupler $(k=c)$, and $E_{Jc}$ is the junction energy of the coupler frequency. The qubit-coupler coupling energy, $E_{kc}$, and qubit-qubit coupling energy, $E_{12}$, are determined by the capacitances between the pads of the floating coupler and the pads of the qubits. The signs of the coupling energies are determined by the capacitance relationships so that $E_{12}$ is always negative, and $E_{1c}$ and $E_{2c}$ have the opposite (same) signs depending on whether the floating coupler has an asymmetric (symmetric) pad configuration, as previously shown in Ref.~\cite{sete2021}. Furthermore, in the derivation of $E_{12}$ in Ref.~\cite{sete2021}, the qubits are assumed to be far enough apart so that the direct qubit-qubit capacitance is negligible. As a result, the magnitude of $E_{12}$ is determined by the capacitances of the nearest-neighbor qubit-coupler pads rather than on an explicit dependence on a direct qubit-qubit capacitance. This makes it possible to design a tunable coupler between qubits with a pitch of more than 2 mm.

The net qubit-qubit coupling is determined by $g_{12}$ and an effective virtual interaction mediated by the floating tunable coupler~\cite{sete2021}: 

\begin{align}
    g &= g_{12} -g_{\rm eff}(\Phi_c), \label{net-g}\\
    g_{\rm eff}(\Phi_c) &= \frac{g_{1c}(\Phi_c)g_{2c}(\Phi_c)}{2}\sum_{k=1}^2\left(\frac{1}{\Delta_k(\Phi_c)} + \frac{1}{\Sigma_k(\Phi_c)}\right),\label{gqq}
\end{align}

where $\Delta_k = \omega_c-\omega_k$ and $\Sigma_k = \omega_c+\omega_k$, and the qubit frequencies are $\omega_{k}$ ($k$=1,2). The coupler frequency is $\omega_c = \sqrt{8E_{\rm Jeff}(\Phi_c)E_{Cc}}-E_{Cc}(1 + \xi/4 )$, where $E_{\rm Jeff}= E_{Jc}\sqrt{1+r^2+2r\cos(2\pi\Phi_c/\Phi_0)}/(1+r)$ \cite{Didier2017} with $r=E_{Jc1}/E_{Jc2}$ and $E_{Jc}=E_{Jc1}+E_{Jc2}$ being the total junction energy of SQUID junctions of the coupler, and $\xi =\sqrt{2E_{Cc}/E_{Jc}}$. When $g_{12}$ is large enough to offset $g_{\rm eff}$, a $g=0$ condition can be reached for specific values of $\omega_c$ which turns off the coupling between the qubits. 

From Ref.~\cite{sete2021}, the direct qubit-qubit coupling $g_{12}$ in Eq.~\ref{g_direct} is not significantly limited by the qubit-qubit separation. As a result, the coupled qubits can be spatially separated onto different chips as long as there is sufficient capacitance between nearest-neighbor qubit-coupler pads. 

\subsection{Multi-Chip Tunable Coupler Assembly} 
To create a tunable coupler transmon that bridges multiple chips, we consider an architecture consisting of two distinct qubit chips bonded to a common carrier chip. Qubit Chip 1 includes one qubit as well as the SQUID loop defining the tunable coupler and the second qubit is placed on a separate Qubit Chip 2. The SQUID loop of the tunable coupler is fabricated on a qubit chip instead of a separate chip in order to reduce the number of different chips requiring complex lithography. To complete the qubit-coupler-qubit unit, the tunable coupler arm extends through the carrier chip as shown in Fig.~\ref{fig:device diagram}(a)-(b). An arbitrary number of additional routing chips can be added along the tunable coupler pathway, depending on the multi-chip assembly needs [Fig.~\ref{fig:device diagram}(c)]. This yields an architecture that is modular and flexible.

\begin{figure*}[!htb]
    \centering
    \includegraphics[width= \linewidth]{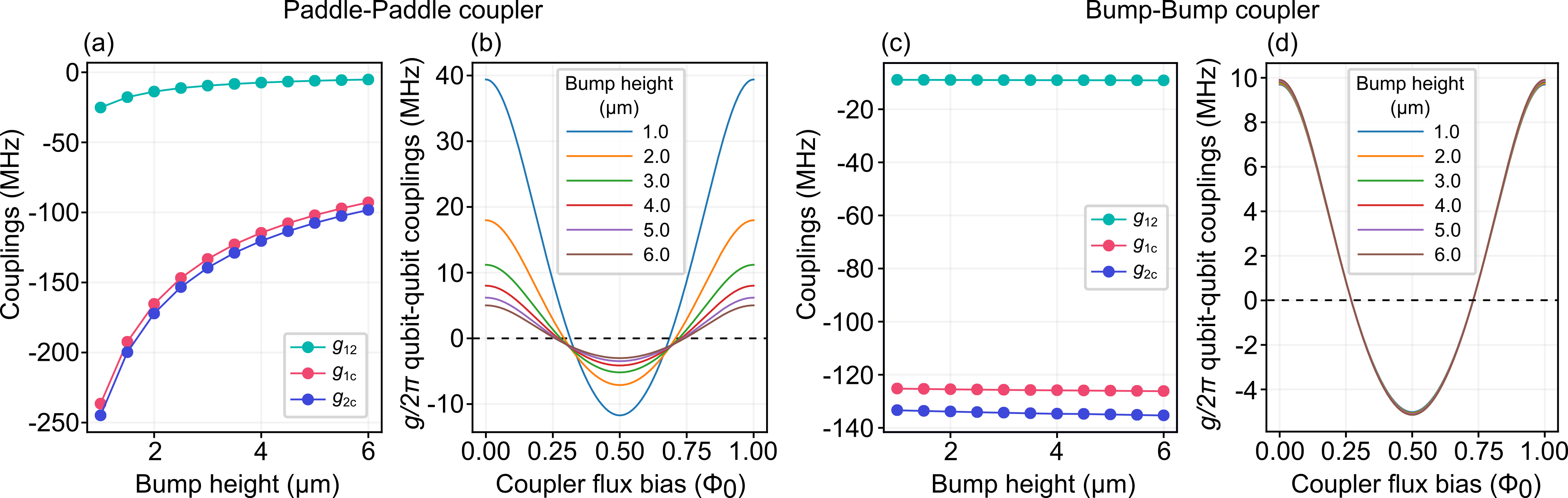}
    \caption{Simulated couplings for a multi-chip coupler assembly with variable indium bump height. We ran a finite-element analysis and lumped circuit simulation to get the various coupling terms $g_{12}$, $g_{1c}$, and $g_{2c}$. The capacitance matrix for the qubit-coupler pairs was extracted for varying bump heights, which was then used as input to a circuit simulator to extract the coupling terms. (a) Direct couplings $g_{12}$, $g_{1c}$, and $g_{2c}$ versus bump height for paddle-paddle coupler design, which is sensitive to the indium bump height variation when the coupler bridges inter-chip air gap capacitors. (b) The net qubit-qubit coupling versus coupler flux bias for various bump heights. Bump height variation impacts the maximum net coupling and the location of the $g=0$ condition.  (c) Direct couplings $g_{12}$, $g_{1c}$, and $g_{2c}$ versus bump height for bump-bump coupler design. The couplings are insensitive to bump height. (d) The corresponding net qubit-qubit coupling versus coupler flux bias for different values of bump height. Here the net qubit-qubit coupling and the locations of the zero-coupling condition are robust against bump height variation. In the experimental devices the indium bump height is 3.5 $\mu$m.
    }
    \label{fig:simulation}
\end{figure*}

We now consider different methods to define the inter-chip connections of a multi-chip assembly and simulate the impact of the inter-chip connections on the effective qubit-qubit couplings. One method would be to use the inter-chip separation to define the capacitance between transmons on separate chips. This was previously studied in Ref.~\cite{gold2021}, which used metal paddles on the qubit and carrier chips to create a vacuum gap capacitor that mediated the static coupling between qubits. The qubit and carrier chips were bonded together by indium bump connectors, which also determined the inter-chip separation and therefore the static coupling. The static coupling was subject to fabrication variations coming from variations in the indium bump heights, but in practice, the bump height variations did not significantly impact two-qubit gate performance in Ref.~\cite{gold2021} due to the two-qubit gate scheme that was used. 

In comparison to a static coupling device, a multi-chip tunable coupler device has a more stringent constraint on capacitance variations since a zero-coupling point is desired. We simulate the impact of bump height variations on the couplings of a multi-chip tunable coupler device by running a finite-element analysis and lumped circuit simulation to get the coupling terms $g_{12}$, $g_{1c}$, $g_{2c}$(Fig.\ref{fig:simulation}(a) and (c)). The coupling terms are then used in Eq.\ref{gqq} to create qubit-qubit coupling plots (Fig.\ref{fig:simulation}(b) and (d)). We provide representative simulations of the two types of inter-chip connections that we consider: paddle and bump inter-chip connections. 

We first consider the impact of capacitive paddles on an example multi-chip tunable coupler that makes use of capacitive paddles on both coupler arms (paddle-paddle coupler). Similar to in Ref.~\cite{gold2021}, the distance between the paddles is defined by the height of a nearby indium bump bond, as illustrated on the vacuum gap capacitor side of Fig.~\ref{fig:device diagram}(a). The vacuum gap capacitor provides the capacitive coupling between qubit and coupler and contributes to the qubit-coupler coupling energies. As a result, fluctuations in the connector bump height result in variations in $g_{12}$, $g_{1c}$, and $g_{2c}$ [Fig.~\ref{fig:simulation}(a)], which impacts the overall $g$ dependence on coupler flux bias [Fig.~\ref{fig:simulation}(b)]. In particular, the flux bias needed to reach $g=0$ varies with indium bump height, and the maximum magnitude of $g/2\pi$ can vary by $\sim$ 20 MHz for 1 $\mu$m of bump height variation. 

To mitigate the impact of the bump height variation on the capacitive parameters of the multi-chip coupler assembly, we consider a second method for the inter-chip connections: replace the capacitive paddles along the coupler arms and instead directly place indium bump connectors along the path of the coupler arm. In this case, the indium bump connectors provide a continuous superconducting path along the multi-chip coupler. By removing the capacitive variation in the coupler path on both arms, the qubit-qubit coupling dependence on coupler flux is not impacted by bump height variations [Fig.~\ref{fig:simulation}(c)-(d)], providing a more robust design for the multi-chip coupler. Based on these simulations, we design and fabricate multi-chip tunable coupler designs that make use of indium bump connections to minimize the impact of indium bump height variations on coupling: Bump-Paddle, Bump-Bump, and Four-Bump. These designs are shown schematically in Fig.~\ref{fig:device diagram}.   

\subsection{Fabrication} \label{fabrication}
The individual qubit chips are fabricated using standard lithographic techniques on high-resistivity silicon substrates. Superconducting circuit components, including the capacitor pads of the transmon qubit structure, ground planes, signal coplanar waveguides, and resonators, are fabricated in niobium metal by deposition, optical lithography, etch, and liftoff steps \cite{nersisyan2019}. The transmon Josephson junctions are fabricated using electron beam lithography with double-angle evaporation of aluminum with in-situ oxidation to form the tunnel barrier.

The qubit chips are then bump bonded to a carrier chip using indium bump connectors, resulting in a multi-chip assembly with a superconducting Faraday cage around each qubit and routing for microwave signal lines. Flip-chip bonding of the carrier and qubit modules is accomplished through the deposition and patterning of indium bumps of height 6.5 $\mu$m and 40 $\mu$m diameter onto the carrier chip. The qubit chips are flipped and aligned to the carrier before thermo-compression bonding, creating a metallic bond between the carrier and qubit chips \cite{nersisyan2019} which goes superconducting below the transition temperature of indium at $T_{C} = 3.4$ K. In the case of the bump-paddle design, the vacuum gap capacitor is formed by two large area superconducting pads separated by a vertical distance of  $\sim$ 3 $\mu$m.
In this experiment, the qubit transmon structures are entirely contained on the individual qubit chips, while the coupler is one resonant superconducting element extending over several surfaces and made up of niobium coplanar transmission lines, aluminum Josephson junctions, and indium bump bonds.

\begin{figure}[tb]
    \centering
    \includegraphics[width=\columnwidth]{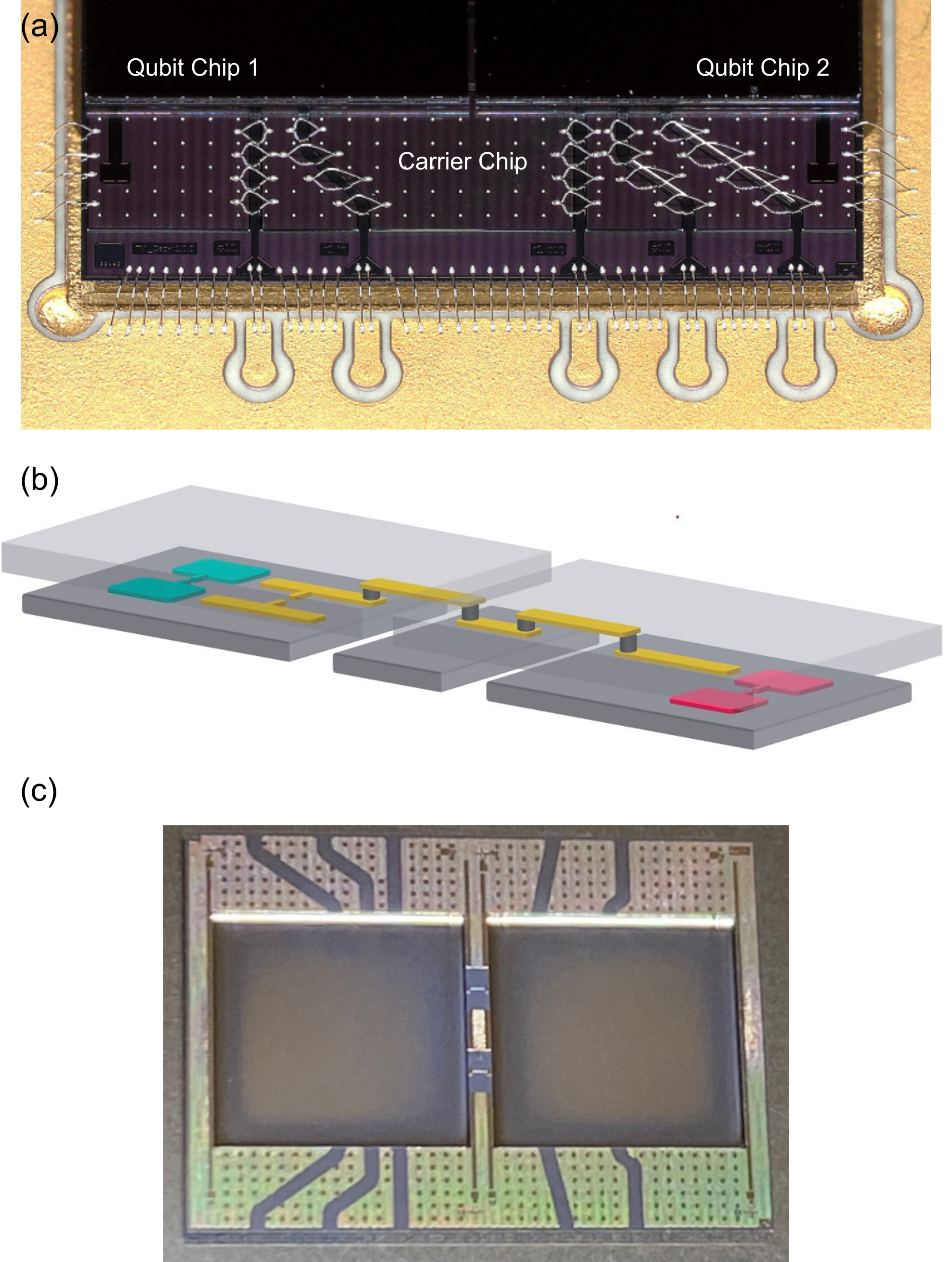}
    \caption{(a) A photomicrograph of an assembled test device with two-qubit chips bonded to a carrier chip. This microscope picture is taken at an angle of 45 degrees with respect to the normal and shows the backs of the two-qubit chips. Control signals and readout lines are connected via microwave lines visible in the foreground on the common substrate chip. (b) A 3D model of the four-bump coupler design (Device C in Fig. \ref{fig:device diagram}(c)) showing the arrangement of four separate indium bump bonds in the coupler arm. (c) A photomicrograph of the Routing Chip designed for this experiment to increase the number of bump bonds in the coupler. The two capped qubit chips sit in the adjacent pockets and bump bonds connect to a signal line on the ridge between as shown schematically in (b).}
    \label{fig:assembly diagram}
\end{figure}

A multi-chip test assembly is constructed through the bonding of two or more chips to a larger carrier chip as shown in Fig.~\ref{fig:assembly diagram}. In this picture, the smaller individual chips have a single flux tunable transmon qubit and corresponding readout resonators, along with part of the tunable coupler oriented along the horizontal axis towards the chip edges that are adjacent to each other. The control lines for flux bias control of the qubits and the tunable couplers, as well as the readout lines, are integrated into the carrier chip. Two qubit chips bonded to a common carrier chip, as shown in Fig.~\ref{fig:assembly diagram}(a), allows the implementation of the Bump-Paddle and Bump-Bump designs. This design could be extended to include an arbitrary number of chips with multi-chip tunable couplers spanning each gap to connect qubits making up a larger processor. A simple version of this is the Four-Bump design shown in Fig.~\ref{fig:assembly diagram}(b) where an additional signal routing chip shown in Fig.~\ref{fig:assembly diagram}(c) is added in.

\section{Experimental demonstration}
We now validate the multi-chip tunable coupler architecture by measuring three devices. On Device A, the inter-chip connections are made with one indium bump connector along the coupler (Bump-Paddle design, Fig.~\ref{fig:device diagram}(a)). On Device B, the inter-chip connections are made with two indium bump connectors (Bump-Bump design, Fig.~\ref{fig:device diagram}(b)). We also test a device with increased qubit-qubit separation on Device C, whose multi-chip coupler goes through an additional routing chip as well as two carrier chips and has a total of four indium bumps (Four-Bump design) as shown in Fig.~\ref{fig:device diagram}(c). To experimentally verify that different types of floating coupler configurations can be used on multi-chip coupler assemblies, we measure devices with both an asymmetric (Devices A and B) and symmetric (Device C) coupler configuration.  

\subsection{Tunable inter-chip coupling}

\begin{figure}[tb]
    \centering
    \includegraphics{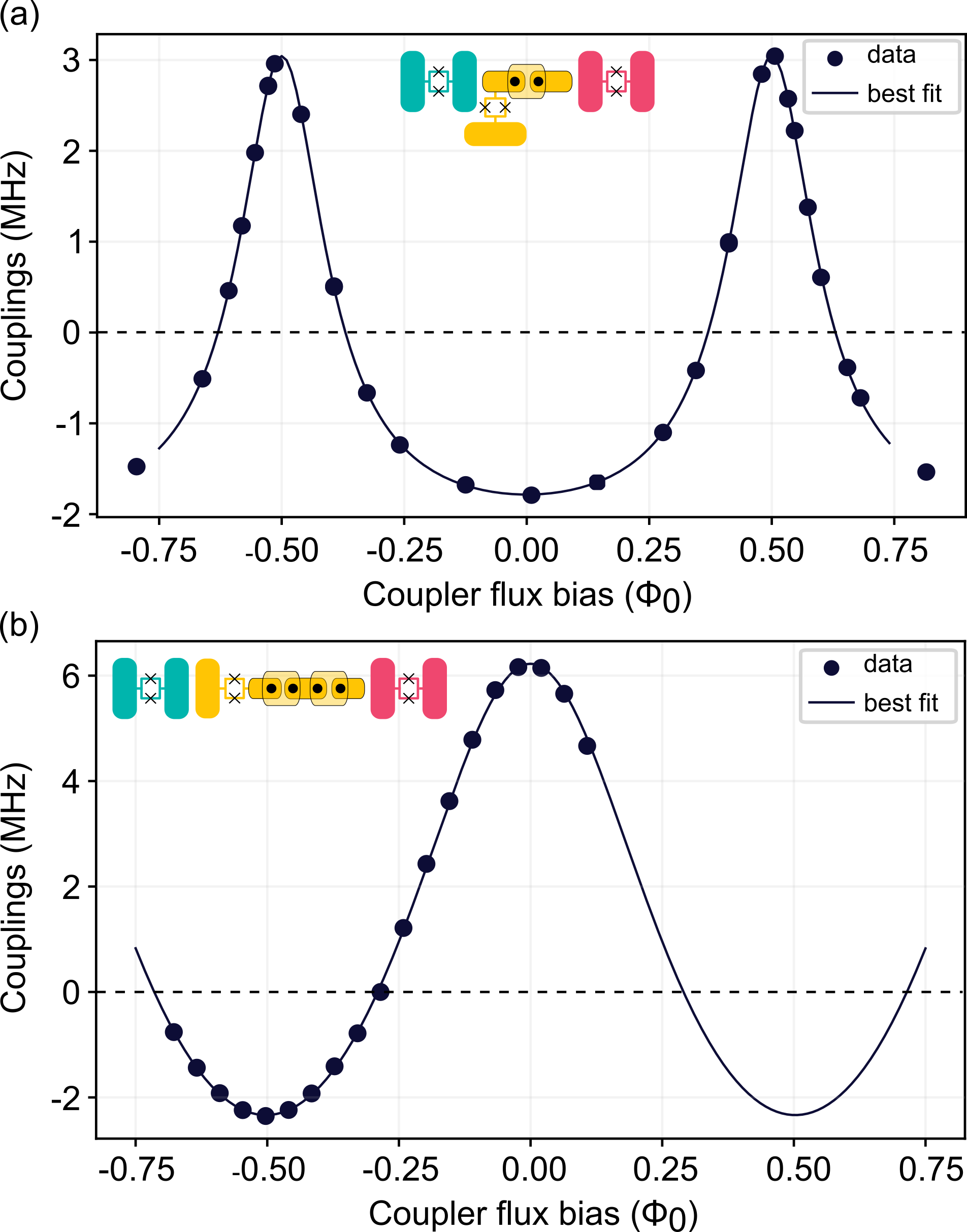}
    \caption{Net qubit-qubit coupling using a tunable coupler connecting two qubits on separate chips. Zero-coupling ($g=0$) is reached for different types of multi-chip coupler designs and for both types of floating coupler configurations. Schematics of the multi-chip coupler designs are included in the insets. (a) Net coupling $g$ vs. coupler flux on Device B, whose multi-chip coupler bridges three chips using the Bump-Bump design. The coupler is in an asymmetric floating coupler configuration. The error bars are smaller than the point size used to plot the graph. (b) Net coupling $g$ measured on Device C, whose multi-chip coupler bridges four separate chips using the Four-Bump design. The net coupling is mediated by a symmetric tunable coupler. The indium bump height is 3.5 $\mu$m. }
    \label{fig:g=0}
\end{figure}

To demonstrate that the multi-chip couplers can mediate both low- and high-coupling interactions between qubits, we measure the net coupling $g$ as a function of coupler flux. The magnitude of $g$ is measured from the energy exchange between the $\ket{10}$ and $\ket{01}$ states, which is enacted using a parametric-resonance gate~\cite{Sete2021-parametric}.  The sign of $g$ is then inferred based on the expected curves for a floating asymmetric or symmetric tunable coupler. Approximate values of the coupling parameters are extracted by fitting the data in Fig.~\ref{fig:g=0} to Eq. \eqref{gqq} ~\cite{sete2021}. In the fitting process, we keep the $E_{Cc}/2\pi$ parameter as a constant taken from design values (113 MHz for the Bump-Bump design and 116 MHz for the Four-Bump design) and allow the other parameters in Eq.~\eqref{gqq} to vary. We show the extracted coupling values and coupler characteristics in Table~\ref{tab:coupling fits}. 

We first present coupling data for a Bump-Bump design (Device B). The tunable coupler has an asymmetric floating coupler configuration so that when the coupler is at its maximum frequency, $g$ will reach its minimum value. The coupling $g/2\pi$ can be tuned from -2 to 3 MHz and a zero coupling condition is attained around $0.37\Phi_0$, which signifies that there was sufficient direct qubit-qubit coupling $g_{12}$ to offset  $g_\text{eff}$ [Fig.~\ref{fig:g=0}(a)]. We extract approximate coupling values $\sqrt{|g_{1c}g_{2c}|}/2\pi\approx93$ MHz and $|g_{12}|/2\pi\approx5$ MHz. These coupling values are similar to those of previous tunable coupler devices fabricated on a single chip~\cite{sete2021}.  

We now look at $g$ when the modular device is extended to four chips with the Four-Bump design on Device C. Device C uses a symmetric tunable coupler, which means that a maximum $g$ is reached when the coupler is at its maximum frequency. For Device C, the multi-chip coupler tunes the value of $g/2\pi$ from -2 MHz to 6 MHz [Fig.~\ref{fig:g=0}(b)]. Despite the more complex multi-chip assembly, $g_{12}$ and $g_{1c}$, $g_{2c}$ on the Four-Bump device are similar to the values extracted from the Bump-Bump device [Table~\ref{tab:coupling fits}]. We also measured similar results for an asymmetric floating coupler with a Four-Bump design.  

\begin{table}[h!]
\centering
\begin{tabular}{ccc}
    \toprule
    & Bump-Bump & Four-Bump \\
    \midrule
    $|g_{12}|/2\pi$ (MHz)            & 5.2 $\pm$ 0.4 & 5.2 $\pm$ 0.9 \\
    $\sqrt{g_{1c}g_{2c}}/2\pi$ (MHz) & 93 $\pm$ 0.9 & 94 $\pm$ 0.6\\
    $E_{Jc}/2\pi$ (GHz) & 58.4 $\pm$ 5.5 & 15.1 $\pm$ 0.2  \\
     $r$& 0.4 $\pm$ 0.02 & 3.0 $\pm$ 0.6\\
    \bottomrule
\end{tabular}
\caption{Coupling values and coupler characteristics of the Bump-Bump design (Device B) and the Four-Bump design (Device C) extracted from fits to measured $g$ data [Fig.~\ref{fig:g=0}(a) and (b)]. To extract the parameters, $E_{Cc}/2\pi$ was taken from the designed value (113 MHz for the Bump-Bump design and 116 MHz for the Four-Bump design) and used as a fixed parameter. The standard deviation for the best-fit value is shown as errors. }
\label{tab:coupling fits}
\end{table}

\subsection{Impact of multi-chip connections on coherence}

\begin{figure}[tb]
    \centering
    \includegraphics{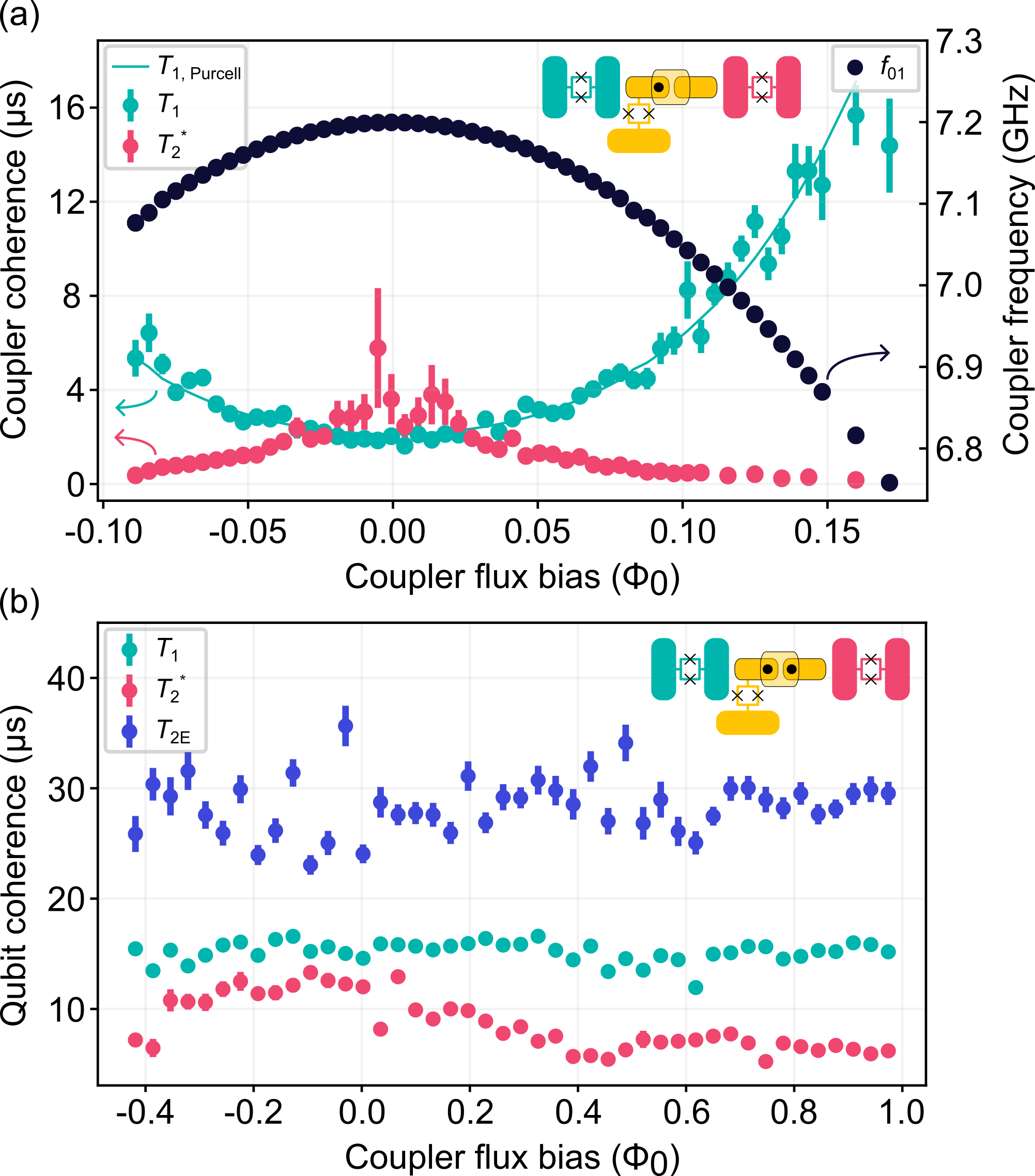}
    \caption{Coherence on devices with indium bump connectors along multi-chip coupler arms. (a) Tunable coupler coherence times of the Bump-Paddle design (Device A). $T_1$ of the coupler follows a Purcell-limited trend. (b) Qubit coherence on the Bump-Bump assembly (Device B) measured as a function of the coupler flux bias. The qubit $T_1$ and $T_{2E}$ do not show any significant degradation as the coupler frequency is lowered towards the qubit frequency. Schematics of the Bump-Paddle and Bump-Bump devices are included in the insets. The indium bump height is 3.5 $\mu$m.}
    \label{fig:coherence}
\end{figure}

Indium bump connectors were placed along the coupler arms to minimize coupling variations coming from bump height variations. We now verify that the indium bump connectors do not significantly degrade the performance of the multi-chip coupler by measuring the coherence times of the coupler on Device A. The coupler frequency $\omega_{c}$ and coherence times near the coupler's maximum frequency $\omega_{c,\text{max}}$ are shown in Fig.~\ref{fig:coherence}(a). 

The coupler is in an asymmetric configuration, so the coupler frequencies are above the qubit frequencies to make a zero coupling condition attainable. At $\omega_{c,\text{max}}$, the coupler frequency approaches the qubit's resonator frequency $\omega_r/2\pi = 7.4$ GHz, and $T_1$ of the coupler decreases. Due to the small detuning between the coupler frequency and qubit resonator frequencies, the $T_1$ values of the coupler are mainly Purcell-limited.  We extract the resonator-coupler coupling strength $g_{rc}$ by fitting the equation for a Purcell-limited $T_{1,\mathrm{Purcell}}$ to the coupler $T_1$ values \cite{Kleppner81,Goy83,Sete14}

\begin{align*}
    T_{1,\mathrm{Purcell}} = \frac{(\omega_c-\omega_r)^2}{\kappa_{r}g_{rc}^2},
\end{align*}

where $\kappa_{r}$ and $\omega_r$ are the decay rates and frequencies for the qubit resonator. The fit yields a coupling value $g_{rc}/2\pi=54$ MHz. However, because Device A uses a Bump-Paddle design instead of a Bump-Bump design, coupling values on the device generally overshot design values due to the extra variability introduced by the additional gap paddle, making it difficult to compare the extracted $g_{rc}$ value with a simulated value. In general, though, $T_1$ of the coupler shows a dependence on coupler-resonator detuning that is consistent with the Purcell effect, which suggests that the quality factor of the multi-chip tunable coupler is not detrimentally impacted by the nearby indium bump connectors. Furthermore, when the coupler was at $\omega_{c,\text{max}}$, $T_1$ of the coupled qubits were around 20 $\mu$s, which suggests that the qubits are not limited by the $<5~\mu$s $T_1$ of the tunable coupler at the coupler's maximum frequency. 

We also measure the qubit coherence as a function of the tunable coupler flux when an additional indium bump is added to the coupler arm on Device B, which has an asymmetric coupler with frequencies above the qubit. The coupler is furthest from the qubit frequency at zero coupler flux bias and approaches the qubit frequencies when the coupler flux bias is at $0.5\Phi_0$. $T_1$ and $T_\text{2E}$ of the qubit do not change in this range as shown in Fig.~\ref{fig:coherence}(b), so the qubit coherence is not limited by the coupler coherence times. There is some degradation in $T^*_2$ in the measurement run, but the degradation is not related to the coupler frequency or the amount of flux applied to the coupler. This degradation is likely coming from the qubit moving off of $\omega_\text{max}$ during the measurement due to classical flux cross-talk that was not corrected. Coherence times $T_1$, $T^*_2$, and $T_{2\text{E}}$ on the other coupled qubit do not show any change with coupler bias. 

\subsection{Two-qubit gate performance}

\begin{figure}[tb]
    \centering
    \includegraphics[width=\linewidth]{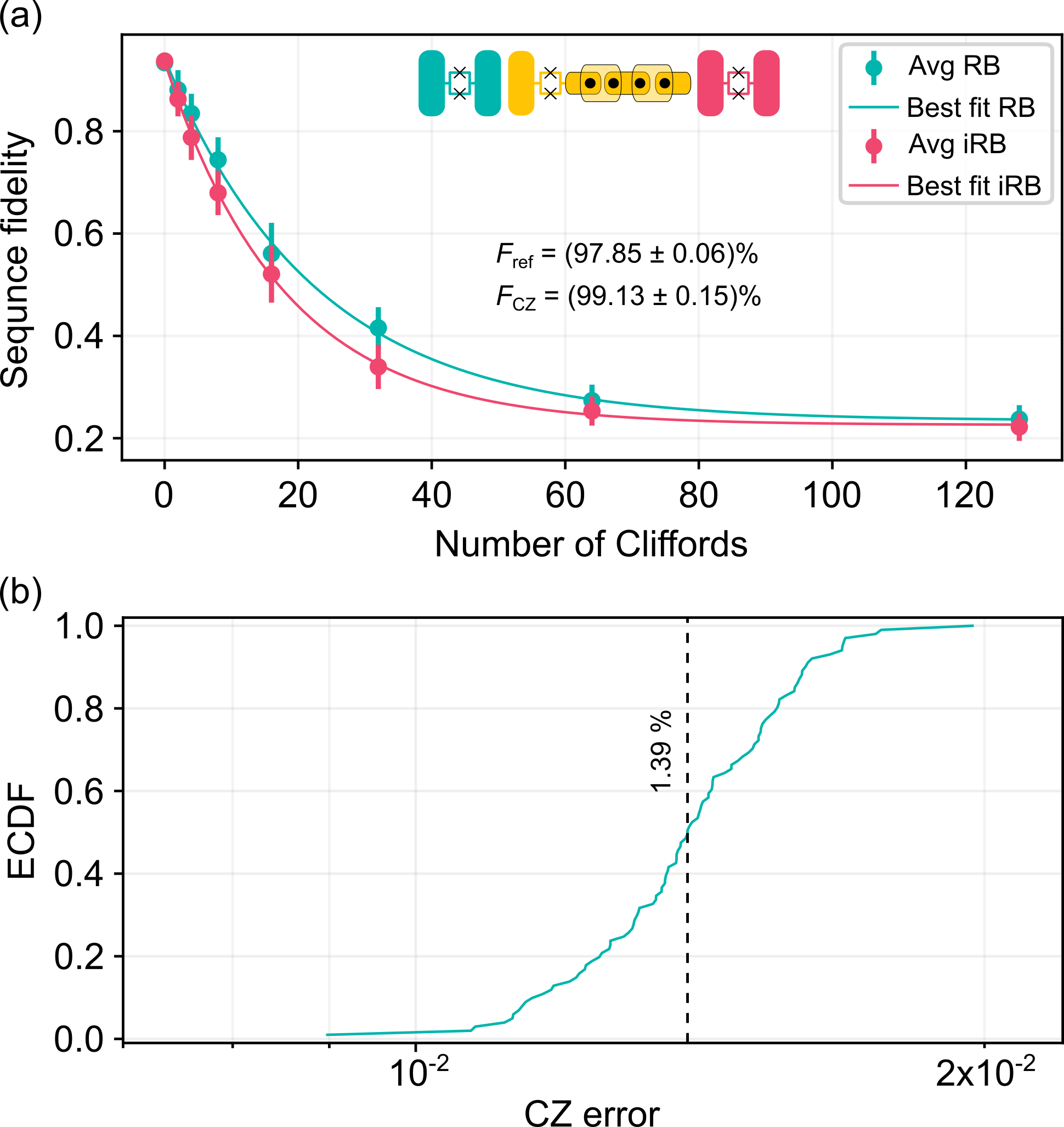}
    \caption{Benchmarking CZ gate fidelity with randomized benchmarking (RB) on a Four-Bump device (Device C). To enact the CZ gate, the multi-chip tunable coupler was biased from zero coupling to maximum coupling. (a) The interleaved RB fidelity is  99.13\%, which is near the coherence-limited value. (b) The ECDF of repeated interleaved RB, with a median error of $1.39\%$. The CZ gate is measured over seven hours without re-calibrations.}
    \label{fig:two-q gate}
\end{figure}

As a final validation of the multi-chip coupler architecture, we use the tunable coupler to entangle two qubits on separate chips. We look specifically at implementing a CZ gate between qubits on Device C, whose coupler bridges four chips and traverses four indium bumps. At the idling bias, the residual ZZ coupling is measured to be 16.5 kHz and the simultaneous single qubit gates have fidelities exceeding $99.7\%$, benchmarked via randomized benchmarking.

Similar to Ref.~\cite{Sete2021-parametric}, the CZ gate is implemented as a parametric-resonance gate, where a modulated flux pulse brings the average frequency of one qubit on resonance with the $\ket{1}\rightarrow\ket{2}$ transition of the other qubit. A dc flux pulse is simultaneously applied to the multi-chip coupler to move $g$ from zero coupling to a larger magnitude. On Device C, we enact a 56 ns CZ gate by moving the multi-chip coupler frequency to the maximum coupling.  

We benchmark the CZ gate using two-qubit interleaved randomized benchmarking (iRB) \cite{Jay12}, shown in Fig.~\ref{fig:two-q gate}(a), and measure the CZ gate fidelity to be $F_{\rm CZ} = 99.13 \pm 0.15 \%$. The coherence-limited fidelity can be estimated using 

\begin{align*}
    \mathcal{F}=& 1 - \frac{1}{5}\left(\frac{1}{T_{1,q1}} +\frac{1}{\tilde T_{1,q2}}\right)t_{\rm pad}
     - \frac{2}{5}\left(\frac{1}{T^*_{2,q1}} + \frac{1}{\tilde T^*_{2,q2}}\right)t_{\rm pad} \notag\\
    &- \frac{19}{160}\left(\frac{1}{T_{1,q2}} + \frac{1}{\tilde T_{1,q2}}\right)t_{\rm flat} - \left(\frac{61}{80}\frac{1}{T^*_{2,q2}}+ \frac{29}{80}\frac{1}{\tilde T^*_{2,q2}}\right)t_{\rm flat},
\end{align*}

where the tilde refers to the coherence time of the modulated qubit. The CZ gate time includes the gate interaction time $t_{\rm flat}$ = 56 ns and the padding on each side of the flux pulses $t_{\rm pad}$ = 4 ns. The measured coherence times are $T_{1,q1}=16.4$ $\mu$s, $T^*_{2,q1}=8.4$ $\mu$s, $\tilde T_{1,q2}=11.5$ $\mu$s, and $\tilde T^*_{2,q2}=5.7$~$\mu$s, so the coherence-limited fidelity is estimated to be $99.1\%$, which is close to the measured fidelity. We also monitor the stability of the CZ gate by measuring the iRB fidelity over a span of 7 hours without re-calibrating the two-qubit gate and qubit readout, as shown in Fig.~\ref{fig:two-q gate}(b). The two-qubit gate is stable over this time with an average fidelity and standard deviation of $98.61\pm 0.18\%$. 

\section{Conclusion}
We have demonstrated an architecture where a tunable coupler spans multiple chips using either vacuum gap capacitors or indium bump bonds in the coupler arms to route the signal via intermediate chips. This architecture allows the combination of the low-error capabilities of a tunable coupler with modular assembly techniques. Furthermore, the coupler design makes use of the fact that the floating coupler does not depend on a direct qubit-qubit capacitance to reach a zero-coupling condition, therefore allowing qubits to be placed on separate chips. We reduce the impact of fabrication variation on the coupling strength and zero $g$ bias condition by introducing indium bump connectors into the arms of the tunable couplers to replace vacuum gap capacitors, which are otherwise sensitive to the separation distance between bonded chips.

We experimentally validate the multi-chip tunable coupler architecture on three different devices, covering configurations where the coupler bridges 3 chips or 4 chips and the coupler is either in an asymmetric or symmetric configuration. Net coupling values comparable to what has been measured in single-chip devices are measured~\cite{Sete2021-parametric}. Furthermore, the coherence times of the multi-chip coupler and the coupled qubits suggest that the location of the indium bump connectors does not significantly degrade the performance of the transmons on the device. We also check that the multi-chip coupler can mediate high-fidelity qubit-qubit interactions: we measure iRB fidelities up to $99.1$\% when performing a 56 ns CZ gate using a coupler that bridges 4 separate chips. Hence, despite the increased assembly complexity, the multi-chip coupler device performs comparably to similar single-chip devices.  

\section*{Acknowledgments}
The authors would like to thank Riccardo Manenti and Beatriz Yankelevich for initial characterizations of multi-chip coupler devices, and Theo Paran for help with resist development in fabricating the routing chip.

\section*{Contributions}
M.F. and A.Q.C contributed equally to this work. M.F. coordinated the experiment and sample preparation. A.Q.C., E.A.S., and S.P. performed the measurements and data analysis. B.S. designed all the devices and ran the simulations to demonstrate the performance with theory input from E.A.S. F.O., K.U., and V.K. created the cleanroom processes used to fabricate the samples and oversaw the sample bonding. S.P., J.M., and A.B. lead the effort. A.Q.C, M.F., E.A.S., and S.P. drafted the manuscript. All authors contributed to the editing of the manuscript.

\clearpage
\bibliography{references.bib}
\end{document}